\documentclass[10pt,amsmath,amssymb,graphicx,nofootinbib]{revtex4} 
\usepackage{bm}
\usepackage{graphics,graphicx,calc,epsfig,pstricks,bbm} 
\usepackage{amsmath,amssymb,amsfonts} 

\newcommand\Tr{{\rm Tr}}

\newcommand{\be}{\begin{equation}}
\newcommand{\ee}{\end{equation}}
\newcommand{\bea}{\begin{eqnarray}}
\newcommand{\eea}{\end{eqnarray}}
\newcommand{\SO}{\mathrm{SO}}

\newcommand{\su}{\mathfrak{su}}
\newcommand{\SU}{\mathrm{SU}}

\newcommand{\cF}{{\mathcal F}}

\newcommand\R{{\mathbb R}}

\newcommand\e{{\mbox{e}}}
\def\extd{\mathrm {d}}

\begin{document}
\title{Anomalous dimension in semiclassical gravity}
\author{Emanuele Alesci}
\email{alesci@theorie3.physik.uni-erlangen.de}
\affiliation{Universit\"at Erlangen, Institut f\"ur Theoretische Physik III, Lehrstuhl f\"ur Quantengravitation\\ Staudtstrasse 7, D-91058 Erlangen, EU}
\author{Michele Arzano}
\email{m.arzano@uu.nl}
\affiliation{Institute for Theoretical Physics,\\ Utrecht University,\\ Leuvenlaan 4, Utrecht 3584 TD, The Netherlands}

\begin{abstract}
\begin{center}
{\bf Abstract}
\end{center}
The description of the phase space of relativistic particles coupled to three-dimensional Einstein gravity requires momenta which are coordinates on a group manifold rather than on ordinary Minkowski space.  The corresponding field theory turns out to be a non-commutative field theory on configuration space and a group field theory on momentum space.  Using basic non-commutative Fourier transform tools we introduce the notion of non-commutative heat-kernel associated with the Laplacian on the non-commutative configuration space.  We show that the spectral dimension associated to the non-commutative heat kernel varies with the scale reaching a non-integer value smaller than three for Planckian diffusion scales.  
\end{abstract}
\maketitle

\section{Introduction}
One of the most pressing questions in various approaches to quantum gravity concerns their semiclassical limit.  Indeed many quantum gravity candidate models \cite{Rovelli:2004tv,Thiemann:2007zz,Oriti:2009zz,Perez:2003vx,Loll:2007rv} are based on pre-geometric, pre-spacetime and discrete structures from which the ordinary description of spacetime in terms of a continuum geometry should emerge \cite{Loll:2007rv,Ambjorn:2004pw,Alesci:2007tx,Alesci:2008gv,Bianchi:2009ri,Barrett:2009mw,Thiemann:2000bw,Giesel:2006uk,Sahlmann:2002qk}.  On the opposite side semiclassical gravity in its simplest realization, quantum fields on curved space-time (see e.g. \cite{Mukhanov:2007zz}), while being very successful both on the theoretical (e.g. the prediction of black hole quantum radiance) and experimental side (the description of the anisotropies in the cosmic microwave background via fluctuations of a quantum field), offers little hints on how its structural properties could be affected by the quantum dynamics of the background.  One notable exception occurs in three dimensions.   As it is well known, in this case, the theory drastically simplifies and its only non-trivial features can appear at a topological level \cite{Deser:1983tn}.  This means that if one wants to couple point particles to the theory these have to be introduced as topological defects or {\it punctures}.  It turns out that the description of the classical phase space of such ``topologically" gravitating particles requires the introduction of momenta which are coordinates on the manifold of the group of local isometries of the theory \cite{Matschull:1997du}.  The group-like nature of momentum has dramatic consequences for the corresponding quantum field theory.  Indeed plane waves can be seen now as group elements and the fields on configuration space form a non-commutative algebra of functions.  Thus semiclassical gravity in three dimension leads to a {\it non-commutative field theory} (NCFT) in which, as we will see in detail below, Newton's constant appears as a parameter which controls the non-commutativity. The remarkable fact is that the same kind of NCFT has been argued to emerge as an effective field theory in the flat-space limit of three dimensional quantum gravity from a spin-foam based approach \cite{Freidel:2005bb,Freidel:2005me} (see also \cite{Noui:2006kv}).  It seems thus that in three dimensions semiclassical gravity is simple but rich enough to provide a natural bridge to the fully blown quantum theory via non-commutative field theory.  In this work we explore further the properties of this NCFT and introduce, via the usual tools of non-commutative Fourier analysis, a notion of non-commutative heat kernel.  As an application we calculate the spectral dimension associated with the non-commutative Laplacian of the model and show that an interesting phenomenon of dimensional flow to values less than 3 takes place as the diffusion parameter approaches Planckian scales.  This is in agreement with results in various quantum gravity scenarios \cite{Ambjorn:2005db,Ambjorn:2004pw,Lauscher:2005qz,Horava:2009if,Carlip:2009kf,Benedetti:2009ge,Sotiriou:2011mu,Sotiriou:2011zn,Modesto:2009qc,Modesto:2008jz} but the surprising fact is that our findings are simply based on the coupling of point particles to gravity and the way the latter affects their phase space.  In particular in our context we {\it do not} consider any quantization of the underlying spacetime and thus the dimensional flow (and the underlying NCFT) should be just seen as inevitable features of (three dimensional) semiclassical gravity.

\section{Curved momentum space from gravity} 
As mentioned above a striking feature of (topological) gravitational interactions of point particles in three dimensions is the way gravity affects the structure of {\it phase space itself}.  Indeed while other interactions simply introduce an interaction potential term in the Hamiltonian describing the system, in three dimensions gravity leads to group valued momenta and a modified symplectic structure \cite{Matschull:1997du}.  Such modifications of the structure of the ordinary phase space disappear when we switch off the gravitational interaction $G\rightarrow 0$ which thus can be seen as a ``deformation parameter". Here we try to describe in the simplest possible way how group valued momenta emerge in this context.\\
We start by considering a relativistic point particle in three dimensional Minkowski space.  The ``extended" phase space (from now on simply ``phase space") $\Upsilon$ of the particle is given by three-positions and three-momenta both coordinates on three dimensional Minkowski space.  From this space we can obtain the ``physical phase space" restricting to a submanifold of $\Upsilon$ determined by the solutions of the equations of motion i.e. straight line geodesics in space-time and on-shell momenta.  For later convenience we note that since three dimensional Minkowski space is isomorphic as a vector space to the three dimensional Lorentz algebra $\mathfrak{sl}(2)$, to each particle we can associate ``Lie algebra-valued" position ${\bf x}= \vec{x}\cdot \vec{\gamma}$ and momentum ${\bf p}= \vec{p}\cdot \vec{\gamma}$ where $\gamma_a$ are 2x2 traceless matrices.  The phase space is thus a {\it vector space} given by the direct product of two copies of $\mathfrak{sl}(2)$ i.e.  $\Upsilon\equiv \mathfrak{sl}(2) \times \mathfrak{sl}(2)\simeq \mathbb{R}^3 \times \mathbb{R}^3$.\\
Let's now switch on gravity.  Since we are in three dimensions the space will still be everywhere flat except at the location of the particle.  We have there a conical singularity and the metric in cylindrical coordinates will be given by
\be
ds^2 = - d\tau^2 + dr^2 + (1-4Gm) r^2 d\varphi^2\, .
\ee
The length of a circular path centered at $r=0$, the location of the particle, divided by its radius will be less than $2\pi$.  The deficit angle is $\alpha=8\pi G m$, proportional to the mass of the particle $m$.  Since the deficit angle cannot be larger than $2\pi$ the mass is bounded from the above $m < \frac{1}{4G}=\frac{M_p}{4}$ with $M_p$ the Planck mass.  We want to characterize the phase space of such a particle and understand what modifications are introduced by the effect of gravity.  In order to do so we first have to remove the point-like singularity from space-time.  This can be done by introducing a cylindrical boundary at $r=0$ by relaxing the condition that points with different angular coordinate $\varphi$ should be identified there.  
Such boundary will represent the worldline of the particle.  In order to determine what three-positions and three-momenta are and compare them with the ordinary flat Minkowski case we map the conical space-time into three-dimensional Minkowski space with a cylindrical boundary and a wedge ``cut-off" representing the deficit angle of the cone.  We thus have a map ${\bf q}$ from the simply connected part of the space-time manifold to $\mathfrak{sl}(2)$ which we identify as a vector space to Minkowski space.  Now, position and momenta are defined with respect to a reference frame.  In Minkowski space all choices of reference frames are equivalent but this is not the case on our conical space-time manifold where we have {\it differently oriented local frames}.  Thus besides ${\bf q}$, which maps the points on conical space to points on Minkowski space, we also need a map of the local reference frames on the conical space-time to the ``background" Minkowski frame.  Such two frames will be in general related by a Lorentz transformation and thus we need an $SL(2)$-valued function ${\bf U}$ in order to specify this information.  The pair $({\bf q},{\bf U})$ provides an isometric embedding of the bundle of local frames on the simply connected part of the manifold into Minkowski space \cite{Matschull:1997du}.  The freedom of choosing the ``background" frame is reflected in the freedom of transforming the functions ${\bf q}$ and ${\bf U}$ via a ``rigid" Poincar\'e transformation
\be
{\bf q}\rightarrow {\bf L}^{-1} ({\bf q}-{\bf n}) {\bf L}\,,\,\,\,\,\,\,\,\,\, {\bf U}\rightarrow {\bf L}^{-1} {\bf U}\, 
\ee
with ${\bf L}\in SL(2)$ and ${\bf n}\in \mathfrak{sl}(2)$ a Lorentz rotation and a translation respectively.  This is an important observation since we need to specify how the values of the embedding functions on the two faces of the wedge ${\bf q}_{\pm}$ and ${\bf U}_{\pm}$ should be identified.  From what we just said in order to have a consistent picture the values of the functions should be related by a generic Poincar\'e transformation i.e.
\be
{\bf q}_+\rightarrow {\bf P}^{-1} ({\bf q}_- -{\bf v}) {\bf P}\,,\,\,\,\,\,\,\,\,\, {\bf U}_+ \rightarrow {\bf P}^{-1} {\bf U}_-\, \,.
\ee
We now have all the ingredients to specify what the phase space of the particle looks like.  The three-position i.e. the location of the worldline in the auxiliary Minkowski space is given by the values of the function ${\bf q}$ on the cylindrical boundary i.e. $\bar{{\bf q}}={\bf q}|_{r=0}$.  
In principle $\bar{{\bf q}}$ is a function of $t$ and $\varphi$, however if we want the cylindrical boundary look like a worldline we must impose the additional condition 
that $\bar{{\bf q}}$ depend only on time\footnote{It can be shown \cite{Matschull:1997du} that such condition is equivalent to impose that the $\varphi$-components of the 
metric vanish on the boundary.} 
$\bar{{\bf q}}\equiv{\bf x}(t)$.
Thus the three-positions of the particle will be still given by a vector, namely the ``coordinates" of an element of $\mathfrak{sl}(2)$.  The situation for three momenta is a little trickier.  First consider the matching condition
\be
\bar{{\bf q}}_+\rightarrow {\bf P}^{-1} (\bar{{\bf q}}_- -{\bf v}) {\bf P}\,.
\ee
Since $\bar{{\bf q}}_+=\bar{{\bf q}}_-={\bf x}(t)$ is the location of the particle, taking the derivative with respect to time of the equation above we obtain that {\it the velocity of the particle must commute with the group element ${\bf P}$} (remember that ${\bf v}$ is constant).  This implies that three-momentum vectors have to be proportional to the {\it projection} of the group element ${\bf P}\in SL(2)$ on its Lie algebra $\mathfrak{sl}(2)$, i.e. if we write ${\bf P}$ in its matrix expansion
\be
{\bf P} = u \mathbbm{1} + 4\pi G \vec{p} \cdot \vec{\gamma}\, 
\ee
we discard the part of ${\bf u}$ proportional to the identity matrix elements and take ${\bf p}=\vec{p} \cdot \vec{\gamma}$.  Notice that now the components of the momentum vector are {\it coordinates on a group manifold}, indeed the condition $\det {\bf U} =1$ implies that 
\be
u^2 - 16\pi^2 G^2 \vec{p}^{\, 2} = 1
\ee
the equation of a hyperboloid embedded in $\mathbb{R}^4$.  The phase space in the presence of topological gravitational ``backreaction" is thus $\Upsilon_G =  \mathfrak{sl}(2) \times SL(2)\simeq \mathbb{R}^3 \times SL(2)$.  We see that the deepest reason why momentum space is curved or ``deformed" is that in order to specify the direction of motion of the particle we have to glue two patches of Minkowski space with a Lorentz transformation.  In Minkowski space we just need vectors to describe position and velocity of the particle, when gravity is switched on we cannot use just vectors but the new picture requires the information contained in a Lorentz transformation.\\
It will be useful before we proceed to make a short digression on the definition of the physical phase space and the associated mass-shell relation. Intuitively in order to determine the mass shell we need to find a characterization of the mass of the particle and relate it to the notion of generalized momentum.  As we saw above the mass of the particle is proportional to the deficit angle of the conical space.  A way to measure the deficit angle is to transport a vector along a closed path around the boundary, as a result it will be rotated by the deficit angle $\alpha= 8 \pi G m$.  The ``holonomy" $\bar{\bf{P}}$, is connected \cite{Matschull:1997du} to the group element ${\bf P}$ by the simple relation
\be
\bf{P} = \bf{U}\, \bar{\bf{P}}\, \bf{U}^{-1}\, .
\ee
This can be interpreted geometrically as follows: consider the ``background" Minkowski frame as the reference frame of an ``observer" sitting at infinity on the conical space-time.  The Lorentz rotation ${\bf P}$ describes the change in the orientation of the reference frame in going across the cut in Minkowski space.  The latter can be also seen as the Lorentz rotation obtained by first transporting the reference frame from infinity to a neighbourhood of the particle, $\bf{U}^{-1}$, then around the particle, the holonomy $\bar{\bf{P}}$, and finally back to infinity $\bf{U}$.\\
Returning back to the mass shell condition we see that the physical momenta will be characterized by all the holonomies which represent a rotation by $\alpha= 8 \pi G m$.  Such requirement imposes the restriction 
\be
\frac{1}{2}\mathrm{Tr} (\bar{{\bf P}}^2) = \cos (4\pi Gm)\,\,\,\longrightarrow \,\,\,\,  \vec{p}^{\, 2} = - \frac{\sin^2(4\pi G m)}{16 \pi^2 G^2}\, ,
\ee
on the ``physical" holonomies.  Classical particles still move along straight lines as in Minkowski space, the non-trivial features emerge at the multiparticle level where the crossing of worldlines is non-trivial due to the conical space-time geometry and in a deformation of the symplectic structure which disappears in the limit $G\rightarrow 0$.\\


\section{Plane waves, Fourier transform and non-commutative space}
We now want to quantize our relativistic particle coupled to gravity in order to obtain a toy model for a {\it self gravitating} quantum field theory. In ordinary Minkowski space the quantum field theory corresponding to a relativistic particle can be constructed from the space of functions on the mass-shell which is isomorphic via Fourier transform to the space of solutions of the Klein-Gordon equation.  In other words one considers functions on momentum space and then imposes a mass shell constraint.  Such space equipped with an appropriate inner product obtained from the Wronskian of the Klein-Gordon equation will give the ``one-particle" Hilbert space of the theory.  We see that in our case, proceeding in the same way, one is naturally led to consider {\it functions on the momentum group manifold}.  Before we proceed any further though, in view of our discussion on the heat kernel and spectral dimension, we switch to the Euclidean.  In this case the ``phase space" of the particle will become $\Upsilon_G =  \mathfrak{su}(2) \times SU(2)$.  We thus see that in the case of a gravitating particle in three dimensions the phase space of the corresponding field theory is given by {\it the space of function on the group manifold $SU(2)$}.  
Since we want to define a heat kernel we need to associate the mass shell discussed in the previous section to (the Fourier transform of) a Laplacian.  While it is straightforward to write down the equation that functions on the mass shell should satisfy we should also make sure that there exists an appropriate generalization of the Fourier transform to functions on a Lie group.  This is indeed the case (see \cite{Freidel:2005bb,Freidel:2005ec,Dupuis:2011fx}).  The basic object we need in order to define a Fourier transform is a {\it plane wave}.  Momenta are now coordinates on $SU(2)$, to fix our conventions we take the Pauli matrices $\sigma_i$ such that $\sigma_i^2=I$  and taking the SU(2) group parametrization
\begin{equation}
\label{g}
{\bf P}(\theta, \hat n) = \cos\theta \, \mathbbm{1}+ i \, \sin\theta \, \hat n\cdot \vec\sigma,
\quad \theta\in[0, \pi], \quad \hat{n}\in S^2.
\end{equation}
we define $\sin\theta \,\hat{n}=\frac{\vec{p}}{\kappa}$ with $\kappa=(4\pi G)^{-1}$, with these  ``cartesian" coordinates we have
\begin{equation}
\label{gp}
{\bf P}(\vec{p}) = \epsilon\, p_0 \, \mathbbm{1}+ i \,  \, \frac{\vec{p}}{\kappa} \cdot \vec\sigma,
\end{equation}
where $p_0=\sqrt{1-\frac{p^2}{\kappa^2}}$ and $\epsilon=\pm1$ if $\theta\in[0, \frac{\pi}{2}]$ or $\theta\in[\frac{\pi}{2},\pi]$.  Plane waves can be written also in terms of a Lie algebra element  ${\bf x}= x^i \sigma_i \in\mathfrak{su}(2)$ as
\be
e_{\bf P}(x) = e^{\frac{i}{2\kappa} \mathrm{Tr} ({\bf x} {\bf P})}=e^{i\vec{p}\cdot\vec{x} }\, .
\ee
with $\vec{p}=\frac{\kappa}{2i} \mathrm{Tr} ({\bf P} \vec{\sigma})$. Note that $\vec{p}\in \mathbb{R}^3$ are the coordinates of the projection of the momentum group element ${\bf P}$ on its Lie algebra and $\vec{x}\in\mathbb{R}^3$ are just coordinates of a 3-vector. 
The main effect of the group structure of momentum space is that the composition of plane waves is {\it non-abelian} indeed we can define a $\star$-product for plane waves
\be
e_{\bf P_1}(x) \star e_{\bf P_2}(x) = e^{\frac{i}{2\kappa} \mathrm{Tr} ({\bf x} {\bf P_1})} \star e^{\frac{i}{2\kappa} \mathrm{Tr} ({\bf x} {\bf P_2})} = e^{\frac{i}{2\kappa} \mathrm{Tr} ({\bf x} {\bf P_1 P_2})}\, ,
\ee
differetiating both sides of this relation and setting the momenta to zero one can easily obtain a non-trivial commutator for the $x$'s which is just the one of $\mathfrak{su}(2)$
\be\label{spinsp}
[x_l, x_m] = i \kappa \epsilon_{lmn} x_n\, ,
\ee
i.e. the ``coordinates" \footnote{We would like to stress once more that such ``space-time" coordinates should not be confused with the coordinates of the particle which form a perfectly commutative algebra but which however obey a non-trivial Poisson bracket.}  $x_i$ can be seen as equipped with a non-commutative algebra structure.\\
To keep things simpler we now restrict to functions on $\SO(3)\simeq\SU(2)/\mathbb{Z}_2$, namely we use  $\vec{p}=\frac{\kappa}{2i} \mathrm{Tr} (|{\bf P}| \vec{\sigma})$ with $|{\bf P}|=\mbox{sign} (\Tr {\bf \,\,P}){\bf P}$.  It is interesting to write down explicitly the non-abelian composition law that momenta inherit from the non trivial group structure  
\be
\vec{p}_{1} \oplus  \vec{p}_{2} =  p_0(\vec{p}_{2})\, \vec{p}_{1} + p_0(\vec{p}_{2})\, \vec{p}_{2} + \frac{1}{\kappa} \vec{p}_{1} \wedge \vec{p}_{2}=
 \vec{p}_{1} + \vec{p}_{2} + \frac{1}{\kappa} \vec{p}_{1} \wedge \vec{p}_{2}+ \mathcal{O}(1/\kappa^2)\, .
\ee
Let us observe that since plane waves are eigenfunctions of translation generators such non-abelian composition of momenta will correspond to what is know as a non-trivial ``coproduct"
\be
\Delta P_a =  P_a\otimes {\bf 1} + {\bf 1}\otimes P_a + \frac{1}{\kappa} \, \epsilon_{abc} P_b \otimes P_c + \mathcal{O}(1/\kappa^2)\, ,
\ee
A non-trivial co-product (or modification of the Leibniz rule) is the ``smoking gun" of {\it quantum deformations} of isometry algebras of ordinary spaces and this is telling us that $P_a$ belong to a non-trivial Hopf algebra or quantum group.  Thus $\frac{1}{\kappa} = 4\pi G$ can be seen as a deformation parameter and in the limit $\kappa\rightarrow \infty$ one re-obtains the usual action of translations as an abelian Lie algebra.
\newline
Now that we have described our plane waves we can introduce a notion of Fourier transform.  Such transform will be a straightforward generalization of the ordinary Fourier transform now mapping the space of functions $L^2(\SO(3), d\mu_{H})$, equipped with the Haar measure $d\mu_{H}$, onto a space $L^2_{\star}(\R^3, d\mu)$ of functions on $\su(2)$, isomorphic as a vector space to $R^3$, equipped with a non-commutative $\star$-product, and the standard Lebesgue measure $d\mu$.  The group Fourier transform \cite{Freidel:2005bb,Freidel:2005ec,Dupuis:2011fx} is given by
\be
\cF(f)(x) = \int \extd \mu_H({\bf P}) f({\bf P})\, \e_{\bf P}(x)\, ,
\ee
which we can write more explicitly in terms of the ``cartesian" coordinates that we have been using for illustrative purposes so far (and which will actually turn out to be useful in the rest of the paper)
\be
\cF(f)(x) = \frac{1}{4\pi\kappa^3}  \int_{|p| \leq \kappa} \frac{\extd^3\vec{p}}{\sqrt{1-\frac{p^2}{\kappa^2}}} f({\bf P}(\vec{p})) \; e^{i \vec{p} \cdot \vec{x}}.
\label{FT2}
\ee
The last thing we should mention before moving to the discussion of the Laplacian is that the space of functions on $\SU(2)/\mathbb{Z}_2$ can be equipped with a hermitian inner product
\be
(f,g)_{G}=\int \extd \mu_H({\bf P})  \overline{f({\bf P})} h({\bf P})\, , 
\ee
which plays a key role in the construction of a QFT Hilbert space and of the two point function \cite{Arzano:2010jw}.\\
 
\section{Non-commutative Laplacian, Green's function and heat kernel}
The mass-shell condition discussed in Section III suggests that in ``cartesian" coordinates the square of momentum vector $\vec{p}$ is the momentum space counterpart of the differential operator of an equation of motion.  Indeed there is a a rather well developed theory of differential calculus on non-commutative spaces and in particular a Laplacian $\Delta_G$ for functions on the ``spin" non-commutative space (\ref{spinsp}) has been introduced in \cite{Batista:2002rq,Majid:2008iz}.  In these works $\Delta_G$ is defined via the ``bi-covariant" differential calculus compatible with the quantum group isometries of the non-commutative space, and it is shown that the plane waves defined above are eigenfunctions of this Laplacian and of the differentials associated with the calculus.  These differentials are mapped via group Fourier transform into the generators of translations $P_a$ and the Laplacian will correspond to the Casimir operator $\mathcal{C}_{G}(P)= P^a P_a$ (which in this particular ``basis" of the enveloping algebra corresponding to our choice of coordinates on the momentum group manifold is just the same as for ordinary 3d Euclidean space).  The action on plane waves is given by 
\be
\Delta_G\, e_{\bf P}(x) \equiv \mathcal{C}_{G}(P)\, e_{\bf P} = \vec{p}^{\,\,2} \, e_{\bf P} \, .
\ee
The Dirac delta function on the group is given by
\be \label{}
\int \extd^3 x \, \e_{\bf P}(x)  := 8\pi \, \delta ({\bf P})
\ee 
and thus plane waves form an orthogonal set of functions with respect to the inner product 
\be \label{}
\int \extd^3 x \, \overline{\e_{\bf P}}(x) \star  \e_{\bf Q}(x)  := 8\pi \, \delta ({\bf P}^{-1}{\bf Q})\, .
\ee 
In order to introduce the discussion on Green's function and heat kernel we reformulate, as customary, the eigenvalue problem in terms of an auxiliary Hilbert space spanned by kets $|x\rangle$ and we seek an operator $\mathcal{O}$ such that
\be
\mathcal{O}\, | \psi_{\bf P}\rangle = \vec{p}^{\,\,2} | \psi_{\bf P}\rangle 
\ee
i.e. with the same eigenvalues as the Casimir above which can be written as
\be\label{opdec}
\mathcal{O} \equiv \int  \extd \mu_H({\bf P})\, \vec{p}^{\,\,2} \, | \psi_{\bf P}\rangle \langle \psi_{\bf P}|
\ee
Using the decomposition of the unity we have
\be
\int  d^3 x' \langle x| \mathcal{O}  |x' \rangle  \psi_{\bf P} (x') = \vec{p}^{\,\,2} \, \psi_{\bf P}(x)
\ee
with $ \psi_{\bf P}(x) = \langle x |  \psi_{\bf P}\rangle$.  We see that the two eigenvalue problems are equivalent if $e_{\bf P}(x)\equiv \psi_{\bf P}(x)$ and 
\be
 \langle x| \mathcal{O}  |x' \rangle = \Delta_G \, \delta (x-x')\, .
\ee
The Green's function for our Laplacian is defined by the usual relation
\be
(\Delta_G+M^2) \, G(x,x') = \delta (x-x')
\ee
so we have that
\be
G(x, x') = (\Delta_G+M^2)^{-1}\,  \delta(x-x') =  \langle x| (\Delta_G+M^2)^{-1} | x' \rangle\, ,
\ee
thus the Green's function is the kernel of the operator $(\Delta_G+M^2)^{-1}$.  Finally using (\ref{opdec}) we can write
\be
G(x, x') =  \int  \extd \mu_H({\bf P})\,  \frac{e_{\bf P}(x)\, \overline{e_{\bf P}}(x')}{\vec{p}^{\,\,2}+M^2}
\ee
which is equivalent to the Euclidean two point function expressed in terms of expectation values on field operators on a one particle Hilbert space.\\ 
To make contact with the heat kernel let us recall the well known operator identity
\be 
\int_0^{\infty}\, ds e^{-s (\Delta_G+M^2)} \equiv (\Delta_G+M^2)^{-1}
\ee
which implies that for $M=0$
\be
G(x, x') = \langle x| \Delta_G^{-1} | x' \rangle = \int_0^{\infty} \, ds\,  \langle x| \ e^{-s \Delta_G} | x' \rangle =  \int_0^{\infty}\, ds\, K(x,x';s)
\ee
where $ K(x,x';s)$ is the kernel of the heat operator $e^{-s \Delta_G}$ and, as for the Green's function, we can express it as an expansion in terms of plane waves
\be
K(x,x';s) = \int  \extd \mu_H({\bf P})\, e^{-s\, \vec{p}^{\,\,2}} e_{\bf P}(x)\, \overline{e_{\bf P}}(x')= \int  \extd \mu_H({\bf P})\, e^{-s\, \mathcal{C}_G(P)} e_{\bf P}(x)\, \overline{e_{\bf P}}(x')\, .
\label{Knc}
\ee
The expression above is the extension of the ordinary heat kernel to non-commutative ``spin" spacetime (\ref{spinsp}).  We would like to stress here that the same straightforward procedure could be applied to other  ``Lie-algebra" non-commutative spacetimes like, e.g., the Euclidean version of $\kappa$-Minkowski space\footnote{In \cite{Benedetti:2008gu} a proposal for the {\it trace} of the heat kernel for $\kappa$-fields was put forward following an analogy with the undeformed case, we would like to point out that the heat kernel defined following our procedure would lead to the same result.}.


\section{Spectral Dimension}
\begin{figure}
\centering
\includegraphics[scale=0.8]{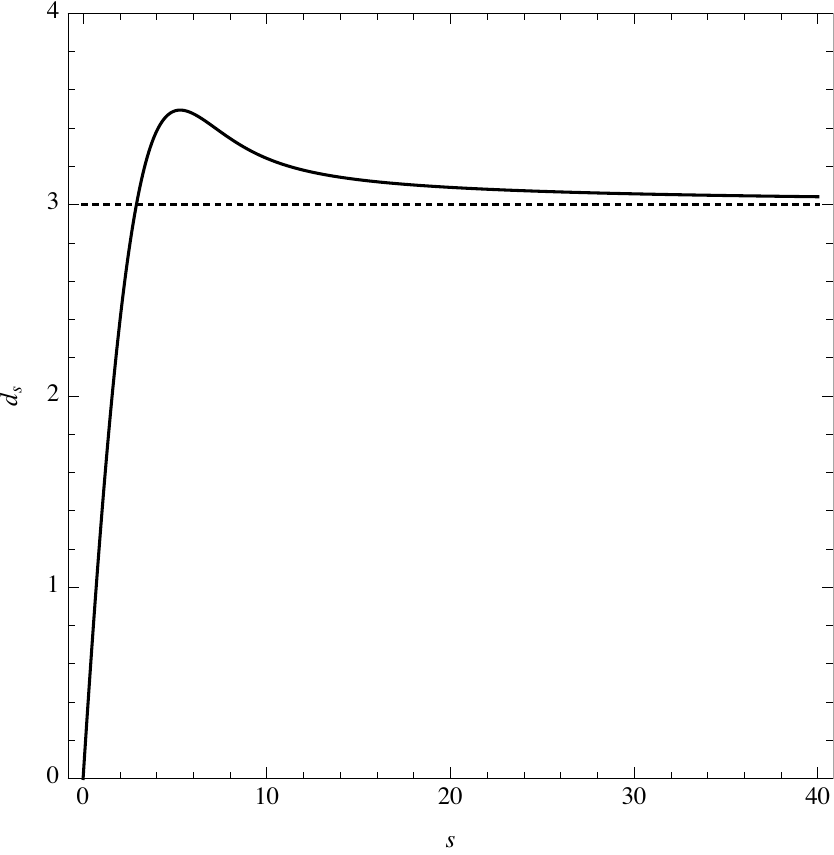}
\caption{Spectral dimension in units $\kappa=1$ (solid line, the dotted is the spectral dimension of ordinary Euclidean space).}
\label{fig1}
\end{figure}

As for an ordinary heat kernel we can define a notion of {\it spectral dimension} for our non-commutative heat kernel
\be \label{d_s}
d_s = -2 \frac{\partial \log \tilde{Tr} K}{\partial \log s}\, ,
\ee
where $\tilde{Tr} K$ is the trace normalized by geometric factors.  Indeed, in ordinary commutative n-dimensional Euclidean space if we take the trace of the heat kernel we find 
\be 
Tr K_{\text{flat}}(x,y;s) \equiv \int d^n x\, K_{\text{flat}}(x,y=x;s) = \frac{V}{4\pi^{n/2}} P_n(s)
\ee
where $V=\int d^n x$ is the volume of space and
\be
P_n(s) = \frac{2}{\Gamma(\frac{n}{2})} \int_0^{\infty} dp\, p^{n-1} e^{-s p^2} \,.
\ee
and thus for the normalized we take $\tilde{Tr}K_{\text{flat}}(s)\equiv P_n(s)$.  For the heat kernel of the ``spin" space (\ref{Knc}), using the Fourier Transform (\ref{FT2}) we have: 
\be
K(x,x';s) =\frac{1}{4\pi\kappa^3}  \int_{|p| \leq \kappa} \frac{\extd^3\vec{p}}{\sqrt{1-\frac{p^2}{\kappa^2}}} e^{-s \vec{p}^{\,2}}\; e^{i \vec{p} \cdot (\vec{x}-\vec{x}')}\, ,
\label{HK transform}
\ee
from which we can easily write down the normalized trace
\be
\tilde{Tr}K_{\kappa}(s)=\frac{1}{4\pi\kappa^3}  \int_{|p| \leq \kappa} \frac{\extd^3\vec{p}}{\sqrt{1-\frac{p^2}{\kappa^2}}} e^{-s \vec{p}^{\, 2}}=
\,\frac{1}{4\pi}\int_{|\mathcal{P}| \leq 1}e^{-s \kappa^2\vec{\mathcal{P}} \,{}^2} \frac{d^3 \vec{\mathcal{P}}}{\sqrt{1-\vec{\mathcal{P}} \,{}^2}}= \int_{0}^{1} e^{-s \kappa^2 \mathcal{P}^2} \frac{\mathcal{P}^2 d\mathcal{P}}{\sqrt{1-\mathcal{P}^2}}
\ee
where we introduced the new integration variable $\vec{\mathcal{P}}=\frac{\vec{p}}{\kappa}$.  The solution of the integral above is known in terms of modified Bessel functions of the first kind $I_n(x)$ (see appendix \ref{BF}) and reads
\be
\int_{0}^{1} e^{-s \kappa^2 \mathcal{P}^2} \frac{\mathcal{P}^2 d\mathcal{P}}{\sqrt{1-\mathcal{P}^2}}=\frac{\pi}{4} e^{-\frac{\kappa^2 s}{2}} \left(I_0\left(\frac{\kappa^2 s}{2}\right)-I_1\left(\frac{\kappa^2 s}{2}\right)\right)\, ,
\ee
so that
\be
\tilde{Tr}K_{\kappa}(s)=\frac{\pi}{4}\,e^{- \frac{\kappa^2 s}{2}} \left(I_0\left(\frac{\kappa^2 s}{2}\right)-I_1\left(\frac{\kappa^2 s}{2}\right)\right)
\ee
The spectral dimension associated to the non-commutative heat kernel can then be easily obtained from the definition \eqref{d_s}  and using the properties of the derivatives of the modified Bessel functions which we recall in Appendix \ref{BF}, we obtain our final result 
\be
d_s=2\left(1+\kappa^2 s+\frac{I_0(\frac{\kappa^2 s}{2})}{I_1(\frac{\kappa^2 s}{2})-I_0(\frac{\kappa^2 s}{2}))}\right)\, .
\ee
In Figure (\ref{fig1}) we plot the spectral dimension in units $\kappa=1$.  
For large values of the diffusion parameter the spectral dimension approaches\footnote{This is easily seen analytically using the the asymptotic expansion \eqref{asym}.} the value $d_s=3$, as expected, since the non-commutativity effects are set by the UV scale $\kappa$ determined by Newton's constant.  However as we decrease the scale at which our particle is probing space we notice a highly non-trivial flow of the spectral dimension. As shown in Figure (\ref{fig1}), as the diffusion time decreases, the spectral dimension increases until reaches the maximum value $d_{s_{max}}\simeq 3.5$ for $s_{max} \simeq 5.3$.  Below $s_{max}$ the spectral dimension starts rapidly decreasing plunging under the ordinary value of 3 at $s\simeq 2.9$.  At Planckian diffusion scales the spectral dimension approaches the value $d_1 \simeq 1.36$ and below this scale the dimension keeps decreasing.  Notice however in this regime our model is no longer reliable since the semiclassical approximation breaks down and one should take into account full quantum gravity effects.
We would like to stress that the running of the spectral dimension is an effect determined solely by the non-commutativity which in turn is related to the group manifold structure of momentum space.  Often in the literature such anomalous behaviour of the spectral dimension is interpreted as a {\it fractal} structure of space \cite{Lauscher:2005qz,Modesto:2009qc,Modesto:2008jz,Modesto:2011,Benedetti:2008gu}.  Indeed it has recently been shown \cite{Arzano:2011yt} that $\kappa$-Minkowski non-commutative space \cite{Majid:1994cy}, with a similar Lie algebra structure as the spin space we considered here, exhibits effective ``fractional" integration measures \cite{Agostini:2006zza} which are directly related to fractal spaces.  Our results thus provide further evidence for the connection between certain classes of non-commutative space-times and multifractional spacetimes \cite{Calcagni:2011kn,Calcagni:2011nc} and their associated field theories \cite{Calcagni:2011sz}.

\section{Discussion}
We discussed how the quantization of relativistic particles coupled to three dimensional Einstein gravity naturally leads to a non-commutative quantum field theory in which the fields are functions of Lie algebra valued ``coordinates" and their Fourier transformed counterparts are function on the corresponding Lie group.  We showed that for the Euclidean counterpart of such field theories it is possible to introduce a notion of non-commutative heat kernel and ``deformed" two-point function.  As an application we calculated the spectral dimension associated with the non-commutative heat kernel which revealed a highly non-trivial behaviour: decreasing the scale at which the system probes the space the spectral dimension grows higher than the usual value of 3 associated to the heat kernel of ordinary Euclidean space.  At 
values of the diffusion scale of few Planck lengths the spectral dimension rapidly decreases to values less than 3.  Such phenomenon of ``dimensional reduction" has appeared in recent years in different quantum gravity scenario, however, what we think is remarkable in the present context is that the same phenomenon appears in a much more ``humble" semiclassical context in which no assumption is made about the structure of ``quantum space-time".  It is simply the ``topological backreaction" of gravity which affects the structure of the phase space of the particle which in turn leads to non-trivial features at the field theory level.  Thus in three dimensions gravity is hinting at a way to go beyond ordinary QFT, introducing a non-commutativity controlled by Newton's constant.  The crucial question to investigate is whether these models will be relevant for four dimensional gravity and in which regimes.

\begin{acknowledgments}
We would like to thank Gianluca Calcagni for enlightening discussions and for helping improve the title of this letter.  M.A. is supported by the EU Marie Curie Actions under a Marie Curie Intra-European Fellowship.
\end{acknowledgments} 

\appendix

\section{Bessel Functions}
\label{BF}

The modified Bessel function of the first kind \cite{Abramowitz} admits the following integral representation for $|\text{arg} \,z|\leq \frac{\pi}{2}$ and $\nu\in \mathbb{R}$
\be
I_{\nu}(z)=\frac{1}{\pi} \int_0^{\pi} e^{z \cos \theta} \cos {\nu\theta} \; d\theta-\;\frac{\sin{\nu\pi}}{\pi}\int_{0}^{\infty} e^{-z\cosh t-\nu t}\;dt
\ee
which simplifies for $\nu=n$ integer to
\be
I_n(z)=\frac{1}{\pi} \int_0^{\pi} e^{z \cos \theta} \cos n\theta \; d\theta
\ee 
A useful formula for the order $k$ derivatives is $I^{k}_{\nu}(z)$ is
\be
I^{k}_{\nu}(z)=\frac{1}{2^k} [I_{\nu-k}(z)+\binom{k}{1} I_{\nu-k+2}(z)+\binom{k}{2} I_{\nu-k+4}(z)\dots+I_{\nu+k}(z)]
\ee
The asymptotic expansion for $\nu$ fixed $|z|$ large and $\mu=4\nu^2$ reads
\be
I_{\nu}(z)\approx \frac{e^{z}}{\sqrt{2\pi z}} [1-\frac{(\mu-1)}{8z}+\frac{(\mu-1)(\mu-9)}{2!(8z)^2})+\dots]\, .
\label{asym}
\ee


\end{document}